\documentclass[10pt,letterpaper]{article}
\usepackage[top=0.85in,left=2.75in,footskip=0.75in]{geometry}

\usepackage{amsmath,amssymb}

\usepackage{subeqnar}
\usepackage{graphicx}
\usepackage{dsfont}
\usepackage{tikz}
\usepackage{booktabs}
\usepackage{tabularx}
\usepackage{comment}
\usepackage{hyperref}
\usepackage{float}

\usepackage{changepage}

\usepackage[utf8x]{inputenc}

\usepackage{textcomp,marvosym}

\usepackage{cite}

\usepackage{nameref,hyperref}

\usepackage[right]{lineno}

\reversemarginpar
\usepackage[colorinlistoftodos]{todonotes}

\usepackage{microtype}
\DisableLigatures[f]{encoding = *, family = * }

\usepackage{array}

\newcolumntype{+}{!{\vrule width 2pt}}

\newlength\savedwidth

\raggedright
\usepackage[aboveskip=1pt,labelfont=bf,labelsep=period,justification=raggedright,singlelinecheck=off]{caption}

\makeatletter
\renewcommand{\@biblabel}[1]{\quad#1.}
\makeatother

\usepackage{lastpage,fancyhdr,graphicx}
\usepackage{epstopdf}
\fancyheadoffset[L]{2.25in}
\fancyfootoffset[L]{2.25in}
\begin{document}
\vspace*{0.2in}


\begin{flushleft}
{\Large
\textbf\newline{Nonlinear control in the nematode \emph{C. elegans}} 
}


Megan Morrison\textsuperscript{1*},
Charles Fieseler\textsuperscript{2},
J. Nathan Kutz\textsuperscript{1},
\\
\bigskip
\textbf{1} Department of Applied Mathematics, University of Washington, Seattle, WA  98195-3925 USA
\\
\textbf{2} Department of Physics, University of Washington, Seattle, WA 98195 USA
\\
\bigskip

* mmtree@uw.edu

\end{flushleft}
\section*{Abstract}
Recent whole-brain calcium imaging recordings of the nematode \textit{C. elegans} have demonstrated that neural activity is dominated by dynamics on a low-dimensional manifold that can be clustered according to behavioral states.  Despite progress in modeling the dynamics with linear or locally linear models, it remains unclear how a single network of neurons can produce the observed features.  In particular, there are multiple clusters, or fixed points, observed in the data which cannot be characterized by a single linear model.  We propose a {\em nonlinear} control model which is global and parameterized by only four free parameters that match the features displayed by the low-dimensional \textit{C. elegans} neural activity.  In addition to reproducing the average probability distribution of the data, long and short time-scale changes in transition statistics can be characterized via changes in a single parameter.  Some of these macro-scale transitions have experimental correlates to single neuro-modulators that seem to act as biological controls,  allowing this model to generate testable hypotheses about the effect of these neuro-modulators on the global dynamics.  The theory provides an elegant characterization of the neuron population dynamics in \textit{C. elegans}.  Moreover, the mathematical structure of the nonlinear control framework provides a paradigm that can be generalized to more complex systems with an arbitrary number of behavioral states.

\section*{Author summary}
\textit{C. elegans} neural activity and its relation to behavior is difficult to characterize as both the dynamics and control are nonlinear. In our work we delineate a set of parsimonious, nonlinear control models that can be minimally parameterized to have the same features as those observed in the neural activity recordings. We analyze the behavior of the models under different parameter regimes and fit a model to \textit{C. elegans} data. Nonlinear interpretable models such as these may give us insight into the control architecture of the \textit{C. elegans} neuron population dynamics, illustrating how intrinsic nonlinearities are exploited for stabilizing and robustly transitioning between behavioral states.


\section*{Introduction}

The emergence of large scale neural recordings across model organisms is revolutionizing the potential for the theoretical modeling of how neuron population dynamics is accomplished.  With the recent advancements in whole brain imaging technologies for the nematode \textit{C. elegans}~\cite{schrodel_brain-wide_2013, prevedel_simultaneous_2014, nguyen_whole-brain_2016}, the relationship between neural activity and behavioral outcomes can be studied in a holistic fashion.   More precisely, \textit{C. elegans} provides a unique opportunity to quantify neuron population dynamics as it has only 302 neurons whose stereotyped electro-physical connectivity map (connectome) is known from serial section electron microscopy \cite{White_paper, chen_wiring_2006}.  We show that the neuron population dynamics of the \textit{C. elegans} nematode can be characterized by a global nonlinear control model which matches experimental measurements.  Moreover, it provides a general mathematical framework that illustrates how nonlinearity can be exploited to produce a global model of neuron population dynamics and how it can be readily applied to more complex model organisms.
\par
Data from \textit{C. elegans} neural recordings show that high-dimensional neuronal activity produces dominant, low-dimensional patterns of activity across the connectome \cite{kato_global_2015,roberts2016stochastic,liu2017functional,kutz2016dynamic,kunert2017multistability,Fieseler_paper1}. 
These low dimensional representations have been considered in posture (behavioral) analysis \cite{stephens2008dimensionality,stephens2011emergence} as well as in the static analysis of calcium imaging data \cite{kato2015global,nichols2017global}.
In previous \textit{C. elegans} modeling work~\cite{costa_adaptive_2019,linderman_hierarchical_2019,nassar_tree-structured_2018,linderman2014discovering, linderman2016recurrent,fieseler_unsupervised_2020}, the dynamics on this low dimensional manifold are described using a set of locally linear models or a controlled linear model.
Respectively, these modeling paradigms can be described with the equations $\dot{\bf x}= {\bf A_i}{\bf x}$, where $i$ refers to multiple segmented state spaces, or $\dot{\bf x}= {\bf A}{\bf x}+{\bf B}{\bf u}$ where ${\bf x}$ is the state space, the dot represents time differentiation and ${\bf u}$ is the control signal. 
The matrices ${\bf A}$ and ${\bf B}$ characterize the intrinsic dynamics, and how actuation forces this dynamics respectively. 
\par
These linear formulations are attractive because of the many theoretical guarantees that exist, including provable control laws~\cite{sontag_mathematical_2013}.   Unfortunately, such a linear control law can only have a single fixed point at the origin.  Thus many nonlinear systems posit this control law near a local fixed point in order to perform control tasks.  This modeling paradigm for control has been exceptionally successful across the engineering, physical and biological sciences.  For data-driven systems, the recently developed {\em dynamic mode decomposition with control} (DMDc) provides a regression method for approximating  ${\bf A}$ and ${\bf B}$ from data alone~\cite{proctor2016dynamic,fieseler_unsupervised_2020,kutz2016dynamic,kaiser_data-driven_2018}.  Alternatively, one can partition the state space behavior into a set of distinct linear models, also known as hybrid or switching dynamical systems, where linear control laws hold in each partition with different matrices ${\bf A_i}$.  This partition is the strategy pursued in recent works~\cite{costa_adaptive_2019,linderman_hierarchical_2019,nassar_tree-structured_2018,linderman2014discovering, linderman2016recurrent}.  In either case, enforcing linearity is highly restrictive, especially when the data suggest that the global dynamics are nonlinear, i.e. multiple fixed points are observed.
Alternatively, if the control signal can be learned from the data, then a single linear model can be used where the control itself effectively accounts for the observed low-dimensional behavior~\cite{fieseler_unsupervised_2020}.    This last modeling effort is a first attempt to model the nonlinear global dynamics within a cohesive, unified (global) framework of linear control but still only produces a single fixed point.  
\par
\begin{figure}[h!]
    \centering
    \includegraphics[width=0.95\columnwidth]{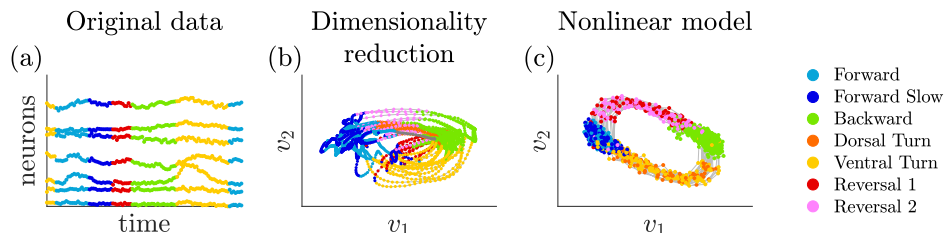}
    \caption{\textit{C. elegans} neural data with a fitted model. (a) Neural activity over time (b) Neural activity in PCA space (c) Dynamical system model fitted to product the same behavioral output in response to \textit{C. elegans} control signals}
    \label{fig:lowD_ex}
\end{figure}

In contrast to linear models which can only support a single fixed point in the dynamics, nonlinear models offer a more flexible architecture for control, especially in systems like the \textit{C. elegans} where multiple behavioral states are clearly observed in the data.  We show that with minimal parametrization, we can construct a global nonlinear model of the underlying \textit{C. elegans} control structure.  Our nonlinear control model removes the need for multiple linear models and provides a parsimonious, global control framework parameterized by only a few parameters and consistent with experimental observations.  Nonlinear control theory takes the form $\dot{\bf x}=f({\bf x}) + g({\bf u})$ where $f(\cdot)$ specifies the nonlinear dynamics and $g(\cdot)$ specifies the actuation on the underlying dynamics.  This provides a theoretical framework for circumventing many of the standard limitations inherited from linear control theory.  This comes at the expense of provable controllability criteria which can be rigorously stated in linear theory.  A fundamental benefit of nonlinear control theory is that one can posit an underlying model with {\em multiple fixed points} where $f({\bf x}_j)=0$ and $j=1,2, \cdots , N$.  In the context of neuron population dynamics and \textit{C. elegans}, these $N$ fixed points correspond to distinct behavioral states, i.e. forward or backward motion.  These multiple and distinct states are clearly observed in the data (See Fig.~\ref{fig:lowD_ex}).  Thus instead of regressing to the matrices ${\bf A}$ and ${\bf B}$ in constructing a linear model, we instead posit a global model whose features are consistent with experimental observations~\cite{kato_global_2015}. 
\par
%
Our model has the flexibility to describe \textit{C. elegans} dynamics under a wide variety of internal states and environmental stimulus. Quantitative work on postural analysis of the behaving \textit{C. elegans} has demonstrated there is low-dimensional structure on the level of individual movements and body bends \cite{stephens2008dimensionality, stephens2011emergence}.
The statistics of how often these movements happen show the presence of a few discrete clusters \cite{pierce1999fundamental, arous2009molecular, wakabayashi2004neurons, churgin2017antagonistic}, or a spectrum \cite{gallagher2013geometry, hums2016regulation} of behavioral strategies that are appropriate in different environments and may even be different between individuals \cite{moy2015computational}.
Recent modeling work has used a conceptual or data-driven model of multiple fixed points in the neuron population phase space \cite{chen2019searching, roberts2016stochastic}.  However, it remains unclear how statistics of transitions between behaviors can be controlled by global parameters, or how individual trajectories through state space are affected in these cases. Our model is able to reproduce the changes in statistics between the large-scale roaming and dwelling behaviors via changing a single global parameter.  In addition, this model reproduces observed short time-scale bursts of reversals interspersed with extremely short-lived forward states. %
\par 
%
Our model further produces testable hypotheses of the effects of neuromodulators on global dynamics. Much work has been done in recent years to extend the understanding of internal \textit{C. elegans} dynamics beyond simple synaptic connections to include additional layers, particularly the slower dynamics of neuromodulators \cite{bentley2016multilayer, komuniecki2014context}. 
Specifically, single molecules and simple neuronal circuits \cite{flavell2013serotonin, arous2009molecular, bhattacharya2014conserved, lim2016neuroendocrine, mccloskey2017food, churgin2017antagonistic, hums2016regulation, wakabayashi2004neurons} have been found to change global statistics related to fundamental behaviors, most clearly the frequency of reversal initiation.
Because our model is able to reproduce macro-scale behavioral changes with a single parameter, we hypothesize that there may be a correspondence between some neuromodulators and our model parameters. 
As we will show, our global nonlinear model is minimally parameterized and provides a parsimonious representation of the neuron population dynamics of the \textit{C. elegans} nematode. These parameters have suggestive connections to experimental work, and some may correspond to one or more neuromodulators. This mathematical framework is general, and can be readily applied to more complex model organisms.

\section*{Results}

We introduce a nonlinear global model for the low dimensional activity of \textit{C. elegans} neuron population dynamics. Any model of this data must satisfy the following requirements:
(1) the general structure of the model must support the two fixed points observed in the data, (2) the model must be flexible enough to accommodate the full range of variability observed in \textit{C. elegans}, and (3) the model must be minimally parameterized such that the modulation of only a few parameters can generate this full range of variability. We start by observing the structure of the data and posit a general model whose parameters can be tuned to generate activity that is analogous to the activity observed in the data. We then explore how experimentally observed changes in \textit{C. elegans} behavior can be explained by the modulation of single parameters. Lastly we observe that the model can be fit to generate the correct activity in response to the control signal derived from the data itself.


\subsection*{A mixture model defines the structure of a nonlinear control model}
\begin{figure}[h!]
        \centering
    \includegraphics[width=0.95\textwidth]{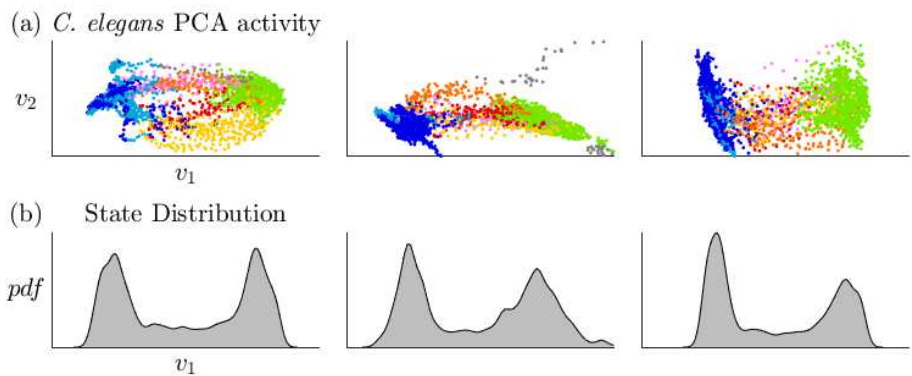}
        \caption{Normalized PCA trajectories of three \textit{C. elegans} paired with probability distribution functions of $v_1$, the dominant mode activity. \textit{C. elegans} have seven behavioral states --- forward (light blue), forward slow (dark blue), dorsal turn (orange), ventral turn (yellow), reversal 1 (red), reversal 2 (pink), and sustained reversal (green). \textit{C. elegans} spend most of their time in a forward or sustained reversal state with irregular transitions between these states.}
        \label{fig:celegans_pdf}
\end{figure}
\begin{figure}[t]
        \centering
    \includegraphics[width=0.95\textwidth]{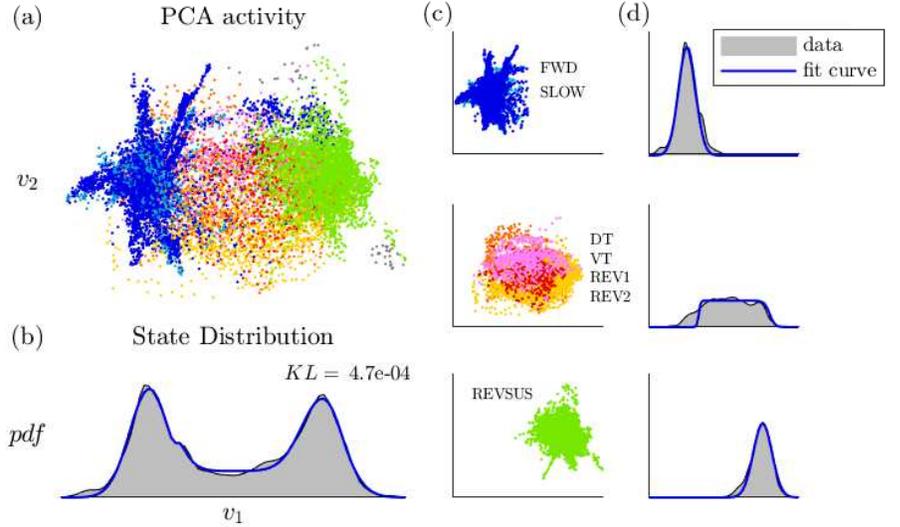}
        \caption{(a) Combined normalized trajectories of five \textit{C. elegans} and (b) the corresponding probability distribution function. (c) The forward (blue) and sustained reversal states (green) can be modeled with a Gaussian mixture model (d) while the transitional states (red, yellow, orange, and pink) can be modeled with a uniform-like distribution constructed with logistic curves.}
        \label{fig:celegan_pdfs_states_all}
\end{figure}

\textit{C. elegans} have been proposed to have seven different behaviors --- forward motion, forward slow, dorsal turn, ventral turn, reversal 1, reversal 2, and sustained reversal~\cite{kato_global_2015}. 
Further references to the Forward behavior denote both the forward motion and forward slow states, and references to the Reversal behavior denote the reversal 1, reversal 2, and sustained reversal states.
We use Calcium imaging data recorded from five different \textit{C. elegans} as their neural patterns express activity corresponding to these different labeled behaviors. We achieve a low-dimensional representation of the activity by performing {\em principal component analysis} (PCA)  on the time series data and focusing on the activity of the dominant PCA modes. Figure~\ref{fig:celegans_pdf} shows the normalized low-dimensional \textit{C. elegans} activity and the probability distribution functions of network states over time in the space of the dominant PCA mode. The \textit{C. elegans} neural network spends the majority of its time in a forward or sustained reversal state with frequent transitions. Figure~\ref{fig:celegan_pdfs_states_all} shows the average dynamics in the dominant feature space for five individual \textit{C. elegans}, with the data labeled according to behavioral responses~\cite{kato_global_2015}. The PCA activity in Fig.~\ref{fig:celegan_pdfs_states_all} has a dominant mode state distribution that is approximated by a three mixture model which is a combination of two Gaussian distributions (forward and backward motion) and a uniform distribution (ventral and dorsal turn transitions). The Kullback-Leibler divergence~\cite{burnham_information_1998} score between the data and three mixture model is $KL = 0.00047$, indicating the proposed Gaussians with uniform mixture model fits the data exceptionally well. The data is decomposed into three constitutive components --- forward motion, backward motion, and turning --- in Fig.~\ref{fig:celegan_pdfs_states_all}(c) while Fig.~\ref{fig:celegan_pdfs_states_all}(d) shows each component's isolated distribution along with the corresponding portion of the fit curve.  
This three-part decomposition of the data {\em probability density function} (PDF) is directly translatable to a nonlinear control framework.
Specifically, two fixed points have been identified, necessitating a cubic dynamical system.
Additional features of the data and how they can be translated into a nonlinear dynamical system are described in Table \ref{table:celegan_features}.



%
\begin{table}[t]
\centering

\begin{tabularx}{\linewidth}{lX}
\multicolumn{1}{c}{\textit{\textbf{C. elegans}}} & \multicolumn{1}{c}{\textbf{Dynamical System}}                            \tabularnewline 
 Two stable fixed points & Globally stable system with 
two sinks                              \tabularnewline \midrule
System functions with variability  & System behavior remains qualitatively constant under small parameter perturbations
\tabularnewline \midrule
Trajectories contain stochasticity & System behavior remains qualitatively constant with the addition of noise  \tabularnewline \midrule
Fixed point locations drift & Behavior remains qualitatively constant despite deformations and shifts to the system
\tabularnewline \midrule
Trajectories tend to follow set paths & System path variability set with damping term
\tabularnewline \bottomrule
\end{tabularx}
\caption{Features exhibited by \textit{C. elegans} neural activity paired with corresponding dynamical system features.}
\label{table:celegan_features}   
\end{table}

\subsection*{Nonlinear global dynamical models for \textit{C. elegans}}

This mixture model suggests a feature space for a model decomposition.  Specifically, it allows us to build a dynamical systems model which accurately reproduces the statistical properties of the global dynamics with minimal parametrization.  The nonlinear parsimonious and global control model takes the form
\begin{subeqnarray}
 &&    x' = y\\
 &&  y'= f(x, \beta) + \gamma y + u(t)
\end{subeqnarray}
where the nonlinear dynamics is prescribed by the cubic
\begin{equation}
  f(x, \beta) = - (x+1) (x-\beta) (x-1)
\end{equation}
which has by construction (for $u=0$) two stable fixed points at $x=\pm1$ and a single unstable fixed point whose location is determined by the parameter $\beta$. 
Additionally, there is damping parameter $\gamma$ and a control input $u(t)$.
These relate to the dominant PCA modes directly, where $x = v_1$ and $y = v_2$

We determine $f(\bf x, \beta)$ by finding a system and control signal that generate the same qualitative attributes as the \textit{C. elegans} PCA data, as outlined in the Methods section. 
Due to the stochastic nature of the observed data, we additionally add stochastic terms and arrive at the system:
%
\begin{align}
\begin{split}
    dx_t &= y_t dt + \sigma dW_t \label{eq:nonlin_model_ep}\\
    dy_t &= -(x_t+1)(x_t-\beta)(x_t-1) dt + \gamma y_t dt + u(t) dt + \sigma dW_t
\end{split}
\end{align}
where $\beta$ and $\gamma$ parameterize the cubic dynamical system, and $\sigma$ and $dW_t$ characterize the Brownian motion which models the noisy fluctuations observed in experiments. 

\begin{figure}[h!]
        \centering
        \begin{tikzpicture}[scale=1]
        \node[inner sep=0pt](russell) at (0,0)
        {\includegraphics[width=0.9\linewidth]{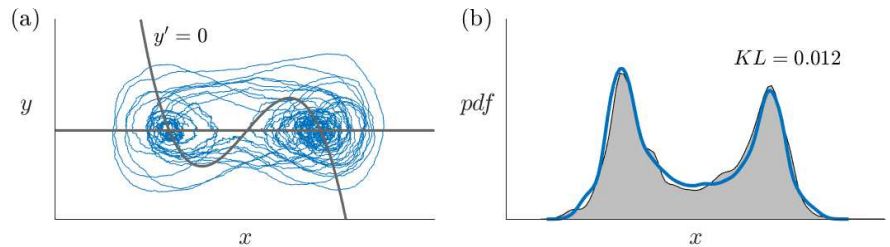}};
        \end{tikzpicture}
        \caption{Fitted stochastic dynamical system with behavior that reproduces the low-dimensional manifold of \textit{C. elegans} neural activity. (a) Trajectories of fitted system $y' = -(x+1)(x-0.03)(x-1) - \frac{1}{2}y + u(t)$, $u = \pm 1$, control signal duration $dur =0.2 + 1.8 \mathcal{U}(0,1)$, control signal frequency $\omega =1/(2.5+\mathcal{U}(0,1))$, and diffusion constant $ \sigma = 0.06$. (b) Probability distribution function of the fitted system (blue) compared with that of the combined normalized \textit{C. elegans} trajectories (grey). The Kullback–Leibler divergence of the two pdfs, $KL = 0.012$, measures the extent to which these two probability distributions differ.}
        \label{fig:model_celegan_compare_pdfs}
\end{figure}

We find these parameter values by fitting the distribution of our model's output to the distribution exhibited by the \textit{C. elegans} data as shown in Figure~\ref{fig:model_celegan_compare_pdfs}. This is a non-convex optimization problem so our method for finding suitable parameter values is to perform a grid search over the parameter space paired with gradient descent. 
While this method finds a suitable collection of parameters, it does not guarantee the optimal solution will be found. We find fitted model parameter values $\beta = 0.03$, $\gamma = -1/2$, $\sigma = 0.06$, and $u(t) = \pm 1$ when $t \in t_{on}$. The duration of the control signal is distributed as $dur = 0.2 + 1.8 \mathcal{U}(0,1)$ while the control signal frequency is distributed like $\omega = 1/(2.5 + \mathcal{U}(0,1))$.

\subsection*{Changes to a single parameter reproduce different long-timescale behaviors of \textit{C. elegans}}

\begin{figure}[!ht]
        \centering
        {\includegraphics[width=0.9\linewidth]{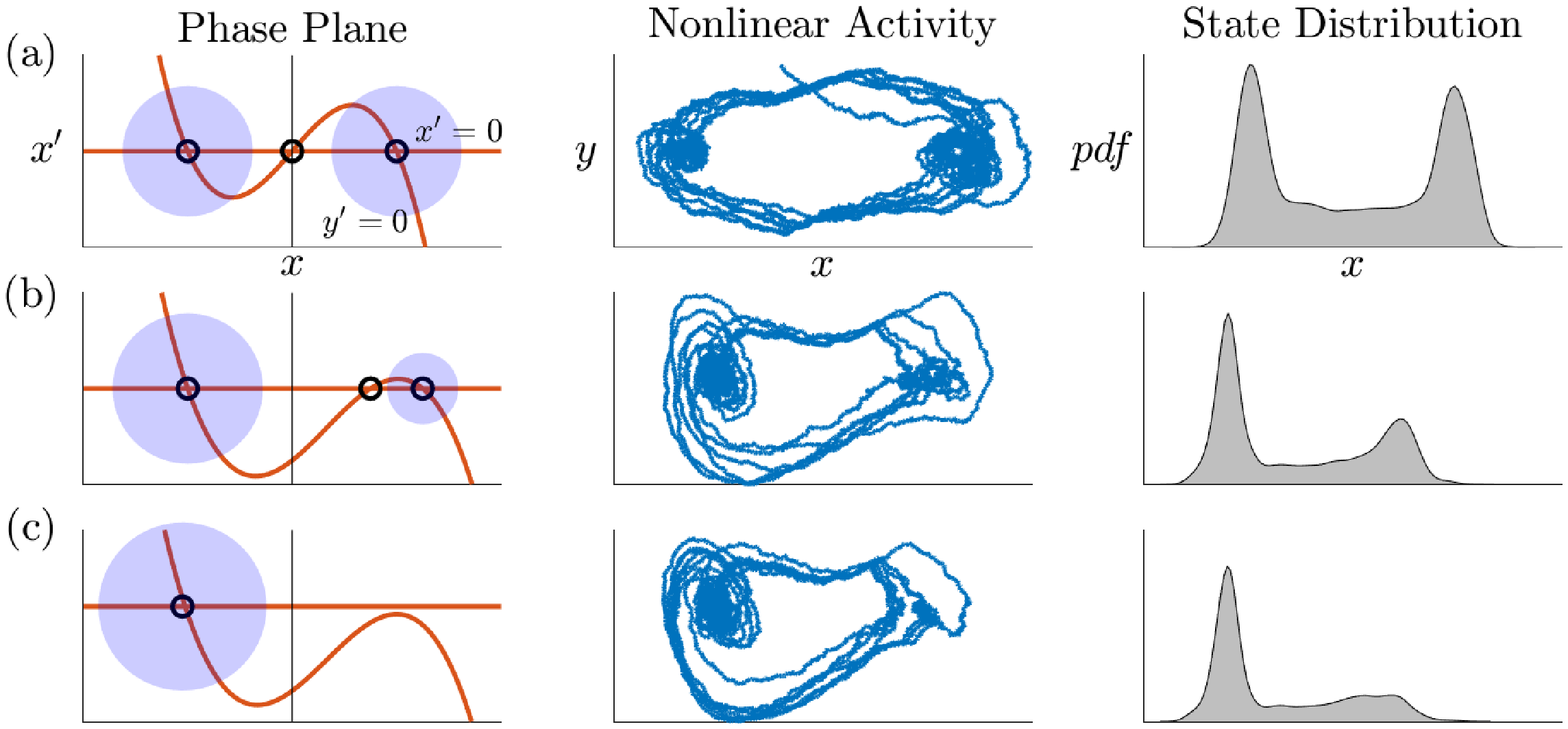}}\\[-.05in]
        {\includegraphics[width=0.9\linewidth]{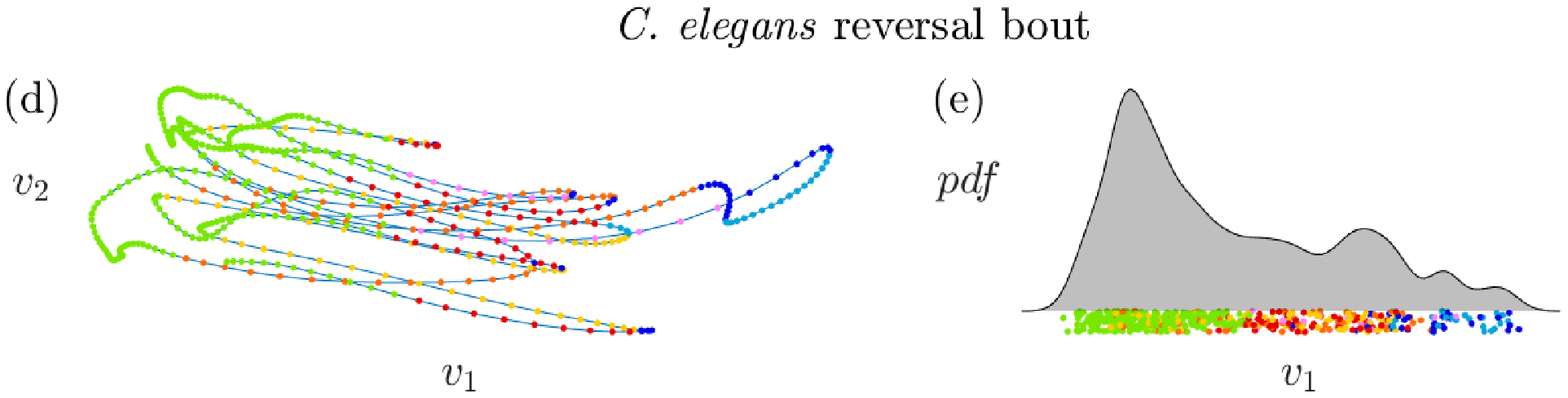}}\\[-.05in]
        {\includegraphics[width=0.9\linewidth]{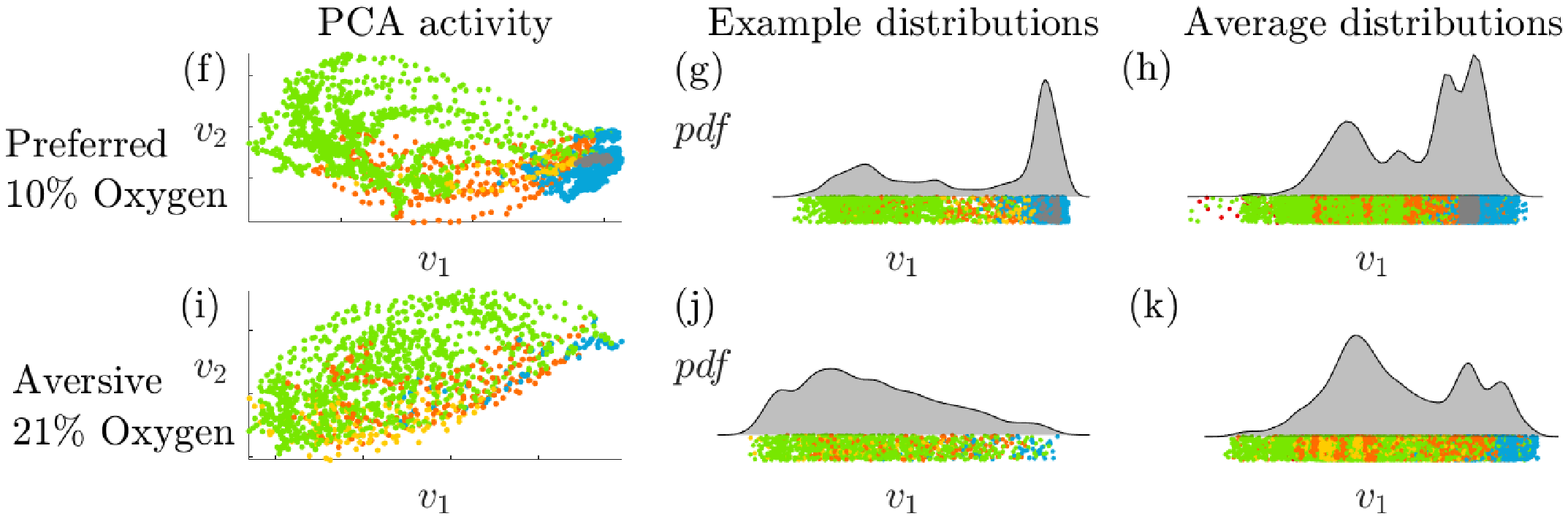}}
        \caption{
        (a) - (c) Phase plane, nonlinear stochastic activity, and state distributions of Eq.(\ref{eq:nonlin_model_ep}) with increasing $\beta$ values. (a) $\beta = 0$ generates equally stable fixed points. (b) $\beta = 0.6$ generates a less stable fixed point which turns into a slow point as the fixed points merge. (c) $\beta, r_2 \in \mathds{C}$ and the right fixed point is lost. (d) \textit{C. elegans} PCA trajectory during a reversal bout and (e) the corresponding distribution. The forward fixed point is unstable during this interval. (f)-(h) \textit{C. elegans} activity in a preferred 10\% oxygen environment which promotes stability in the forward state compared with (i)-(k) \textit{C. elegans} activity in an aversive 21\% oxygen environment which destabilizes the forward state. (f)-(g) PCA activity and distribution of a single \textit{C. elegans} in the preferred oxygen environment compared with the activity of this same \textit{C. elegans} in the aversive oxygen environment (i)-(j). Average distribution for 10 \textit{C. elegans} in the preferred environment (h) compared to the aversive environment (k).}
        \label{fig:models_figs}
\end{figure}

As shown in the methods section, this global model has three fixed points whose stability is determined by the parameter $\beta \in (-1,1)$.  The parameter $\gamma$ determines the linear growth/decay rate of each fixed point.  The parameter 
$\sigma$ controls the amount of stochasticity in the system.  All three parameters are estimated from the data in Figure~\ref{fig:celegan_pdfs_states_all}.
Figure \ref{fig:models_figs}(a)-(c) shows the behavior of Eq.(\ref{eq:nonlin_model_ep}) as a function of $\beta$.  For $\beta=0$, there is a symmetry between the two stable fixed states corresponding to forward and backward motion, which reproduces the long time-scale distribution of behaviors across individuals.  

The statistics of reversal length and frequency change drastically across multiple timescales during the life of a \textit{C. elegans}.
Our nonlinear control model is able to reproduce three very distinct changes in state distribution and switching frequencies seen in these experimental studies via modulation of a single parameters.

The first well-studied change in these dynamics is the switch between dwelling and roaming states \cite{flavell2013serotonin, pierce1999fundamental, arous2009molecular, wakabayashi2004neurons}.
Specifically, the frequency of reversals is much lower in the roaming state, which facilitates the exploration of a larger geographical area. We are able to reproduce this long-timescale behavioral change in our model by reducing the parameter $\omega$ which controls the frequency of the stochastic control signal. 
Several neuromodulators \cite{flavell2013serotonin} and individual neurons \cite{wakabayashi2004neurons} have been implicated in this behavioral change, and thus this model parameter may directly correspond to some function of these chemicals or neuron activity levels.

Two additional behaviors that are not known to be related may in fact operate according to a similar mechanism: reversal bouts, and an increase in reversals in an aversive oxygen environment. 
The reversal bout behaviors, as shown in Figure \ref{fig:models_figs}(d)-(e), are long-lived behaviors that begin in a reversal state, move into a forward motion state but then fail, and return to a reversal state several times in succession. 
This can be clearly related to a change in the parameter $\beta$, which controls the stability of the fixed points corresponding to forward and backward motion.
A known method for experimentally destabilizing the forward state in \textit{C. elegans} is through a modification of their environment. In an environment with a preferred oxygen level of $10\%$, \textit{C. elegans} tend to have stable forward swimming behavior, Figure \ref{fig:models_figs}(f)-(h). When the oxygen in their environment is increases to $21\%$, they exhibit more transient forward swimming behavior, Figure \ref{fig:models_figs}(i)-(k), similar to the observed ``reversal bouts". 
\par
Increasing $\beta$, as shown in Figure \ref{fig:models_figs}(b)-(c), reproduces this unstable forward behavior by retaining the stochastic control signals that would normally transition the system to a forward motion state, but by reducing the stability of that fixed point so that the neural trajectory immediately falls off and returns to a reversal state.
We hypothesize that $\beta$, like $\omega$, also has a biologically correlated neuromodulator or set of neuromodulators and that stabilization of this modulation system would remove the reversal bout phenomenon.
An additional testable prediction is that some subset of neurons correlated with forward motion (e.g. the AVB and RIB pairs) or the ending of reversals (e.g. the SMDD, SMDV, and RIV pairs) may be responsible for stabilizing the forward state and others may be key for initializing the state. 
Opto-genetic manipulation of the ``initiating" neurons without the ``stabilizing" neurons should simply produce a failed forward initialization, as seen in the natural reversal bout.
Similarly, inhibition of the stabilizing neurons should make forward motion an inaccessible state. 
\subsection*{Robustness of results to parameter variations}

\begin{figure}[!ht]
        \centering
        \begin{tikzpicture}[scale=1]
        \node[inner sep=0pt](russell) at (0,0)
        {\includegraphics[width=0.9\linewidth]{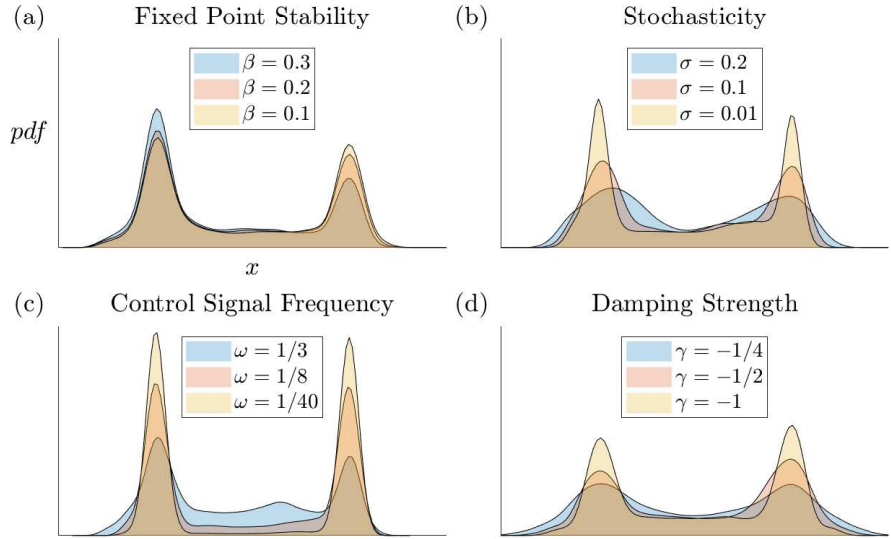}};
        \end{tikzpicture}
        \caption{State distributions of nonlinear models for various parameter regimes. (a) Fixed point relative locations affects their stability. (b) Increasing levels of Brownian motion $(\sigma)$ increases the variation about the fixed points. (c) More frequent control signals more evenly distributes the time spent in stable versus transitional states. (d) Stronger damping in the system keeps trajectories close to fixed points.}
        \label{fig:model_vary_params}
\end{figure}

We now observe how modifying other system parameters affect the state distribution of the nonlinear system's activity. In Figure \ref{fig:model_vary_params}(a) we vary the right fixed point's region of stability by moving the location of the middle fixed point $\beta$. We observe the system spends less time at the right fixed point with a smaller stability region.  
In Figure \ref{fig:model_vary_params}(b) we increase the level of Brownian motion $(\sigma)$ in the system and observe the variability increases in the distributions as a result. In Figure \ref{fig:model_vary_params}(c) we observe that increasing the control signal frequency increases the amount of time spend in a transitional state. Figure \ref{fig:model_vary_params}(d) shows that increasing the damping strength decreases the distribution variability. Observing these parameter variations holistically, we see that the nonlinear model is able to perform the task of switching between fixed points under a wide range of parameter values which insures the integrity of the system and indicates that \textit{C. elegans} dynamics, if comparable to this model, should be able to operate robustly and stably under a diverse array of environments and internal states. 



\subsection*{Nonlinear model recreates dynamical behavior from control signal in the data} 

We now take an alternative approach to fitting a nonlinear model to the \textit{C. elegans} data. Instead of generating a model that creates the correct distribution in response to randomly generated control signals as shown in Figure~\ref{fig:model_celegan_compare_pdfs}, we instead find a model that produces the correct low dimensional activity in response to signals measured from the \textit{C. elegans} neural activity directly~\cite{fieseler_unsupervised_2020}. With this approach we can not only compare state distributions, but can also reconstruct and compare neural activity in the original high dimensional space. After fitting the low-dimensional models to respond correctly to \textit{C. elegans} control signals, we reconstruct individual neuron trajectories using our dominant PCA modes. 

\begin{figure}[!ht]
    \centering
    \includegraphics[width=0.9\columnwidth]{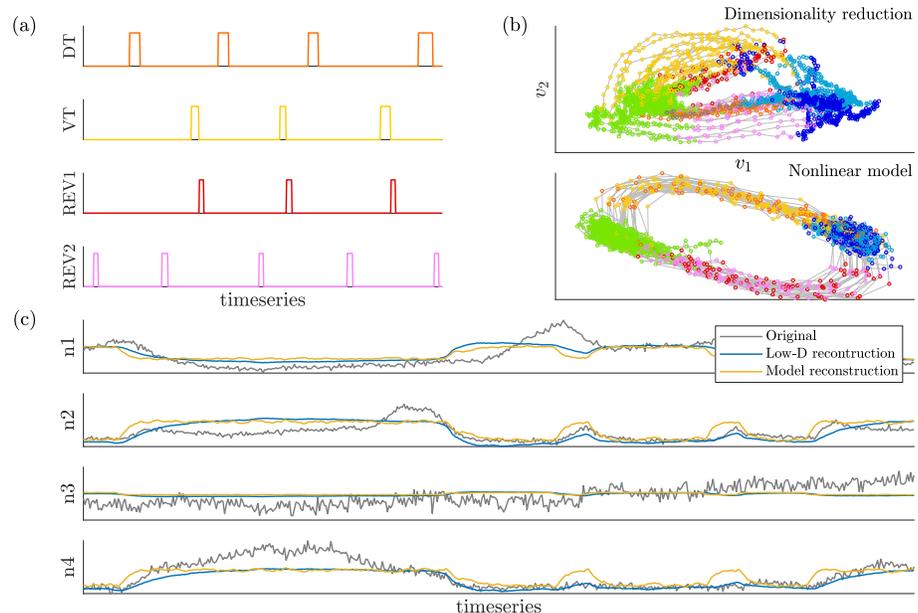}
    \caption{Model is controlled by signals in data and used to reconstruct original neural activity. (a) Four transition signal types --- DT, VT, REV1, and REV2 --- shift \textit{C. elegans} between forward and backward behaviors. (b) PCA space neural activity and model activity controlled by data control signals. (c) Reconstructed activity of four neurons. The nonlinear model reconstruction (yellow) approximates the low-dimensional reconstruction (blue) yet with an enhanced response to changes in activity.
    Both the model and data reconstructions only capture the first-order activity of the original trajectories (grey) due to the information lost in the later modes.}
    \label{fig:reconstruction}
\end{figure}

Figure \ref{fig:reconstruction}(a) shows a timeseries of the four behavioral control signals --- dorsal turn (DT), ventral turn (VT), reversal 1 (REV1), and reversal 2 (REV2). Figure \ref{fig:reconstruction}(b) shows the \textit{C. elegans} neural activity data in PCA space colored by behavioral state along with a fitted model controlled by the same \textit{C. elegans} control signal and colored by this same timeseries of behavioral states found in the data. The forward (blue) and backward (green) timepoints are clustered together with the transition points spanning the path between meaning that the model is able to transition to the correct location in PCA space for each behavioral regime. This model was fit using a grid search of the parameter space. Figure \ref{fig:reconstruction}(c) shows the timeseries reconstruction of four neurons. The model reconstruction fits the low-dimensional reconstruction well, however, neither adequately represents the original timeseries as the first two modes in PCA space only contain a moderate amount of the total variance (e.g. 22\%). This model captures the first-order structures in the system and can certainly be improved by using more PCA modes.  

\section*{Discussion}

We have produced the first global, nonlinear model that can capture the dominant features of low-dimensional neural data.
The model incorporates a stochastic control signal, similar to previous work on stochastic switching models, but extends previous work by explaining incomplete or unsuccessful switching seen in reversal bouts as a change in the stability of the underlying fixed point.
This model is minimally parameterized and changes in several parameters can reproduce changes in behavioral distributions akin to that of known neuro-modulators, thus producing a unifying framework for analyzing various changes in distributions of behavior at multiple timescales.
%
In addition, the framework for building this model can be extended to other complex systems with more behavioral states which are defined by fixed points.

Several modeling strategies have been used to model \textit{C. elegans} behavioral and neural dynamics, and they can be classified in two ways: direct models of the trajectories in neuron space \cite{linderman2014discovering, fieseler_unsupervised_2020, linderman2016recurrent, linderman_hierarchical_2019}, and abstract Markov models \cite{roberts2016stochastic}.
The former has the advantage of describing neuron-level dynamics at the cost of many parameters, generally hundreds.
On the other hand, Markov models do not make specific predictions about neurons or trajectories on the low-dimensional manifold, but generally have a small number of very interpretable parameters.
Our model combines the strengths of both approaches, producing a model of dynamics that is both directly connected to neural activity and has only 4 parameters.
It is unclear if these parameters have biological correlates, but the fact that modulating them produces known behavioral outcomes suggests areas for future experimental work.

This modeling strategy has a few limitations.
In particular, the entire model was constructed and fit using the first two PCA modes, which only account for~22\% of the variance in the data.  Despite this, it provides a model that agrees remarkably well with experimental observations.  Regardless,
it is almost certainly true that important activity is contained in higher PCA modes, particularly when trying to incorporate more complex behaviors.
In addition, it is unclear that PCA modes are the correct basis for producing models whose behaviors have biological correlates.
Work regarding an interpretable choice of basis is ongoing, with nonlinear embeddings offering more flexible possibilities~\cite{lusch_deep_2018,champion_data-driven_2019}. 

Connected to this issue, the model does not clearly differentiate between ventral and dorsal turns.
These behaviors are difficult to clearly separate in the first two PCA modes, even though they are clearly mutually exclusive at the level of muscle activation.
In addition, the individual trajectories were considered stochastic and thus the probability density functions were matched, instead of direct trajectory matching.
Extending our framework to incorporate more subtle and complex behaviors is the subject of ongoing work


The modeling strategy proposed in this paper used polynomials to design fixed points and the transitions between them.
Even if the ``true" function form is more complex, polynomials can be considered a Taylor expansion approximation of those dynamics.
However, no attempt was made to explicitly derive this functional form from neuron-level nonlinearities, or to include information from the known connectome \cite{White_paper}.
A derivation from first principles would be an exciting advance and we hope that our model, as one possible macro-scale model, can facilitate this type of theoretical development.


%




\section*{Methods}

We present a general nonlinear model that can be tailored to fit the features of a certain class of data. More specifically, our general model is capable of describing datasets in which the system transitions between multiple fixed points. The effect of each parameter on the system's behavior is straightforward and the locations and strengths of fixed points and low-dimensional manifolds can be easily determined. Lastly we determine the conditions necessary to make nonlinear control of such systems a possibility.

\subsection*{Nonlinear Dynamical Systems}

Nonlinear dynamical systems are ubiquitous in the engineering, physical and biological sciences for describing many complex phenomenon observed in a diverse number of settings.  Often, simple qualitative models with polynomial nonlinearities are capable of providing remarkable insight into dynamical behaviors.  The nonlinear pendulum, for instance, can be approximate by a Taylor series expansion to characterize the effects of frequency shifts and harmonic generation that is observed in practice.  Inspired by well-studied nonlinearities, we consider dynamical systems of the general form 
%
\begin{equation}
  \dot{\bf x} = f({\bf x},\beta,\gamma) + {\bf B}   u(t)
  \label{eq:nonlinear_control}
\end{equation}

We restrict our focus to polynomial equations with fixed points that can be determined analytically:
\begin{align}
    x' &= y\\
    y'&= f(x) + \gamma y + u(t)\\
    f(x) &= a \prod_{i = 1}^n (x-r_i)
\end{align}
where $f(x)$ is a polynomial with a leading coefficient $a$ and roots $r_i$ and $\gamma$ is the damping parameter. This is a second order nonlinear differential equation which can be expressed as $x'' -f(x)-\gamma x' = 0$. If $\gamma = 0$, the system is undamped and the differential equation becomes $x'' = f(x)$ which has an analytical solution. Often however, the solutions are exceedingly complex and it is preferable to take a qualitative approach.   We choose a system of this form as the fixed points can be easily placed and assigned a stability type (e.g. saddles, sources, sinks, or centers) through parameter selection. All fixed points lie on the x-axis and are placed and manipulated by varying our polynomial roots $r_i$, while fixed point stability types are assigned by manipulating $\gamma$ and $a$ for a given set of roots $r_i$.\\[.2in]


\subsection*{Damping parameter and manifold formation}

\begin{figure}[!ht]
    \centering
    \includegraphics[width=0.9\columnwidth]{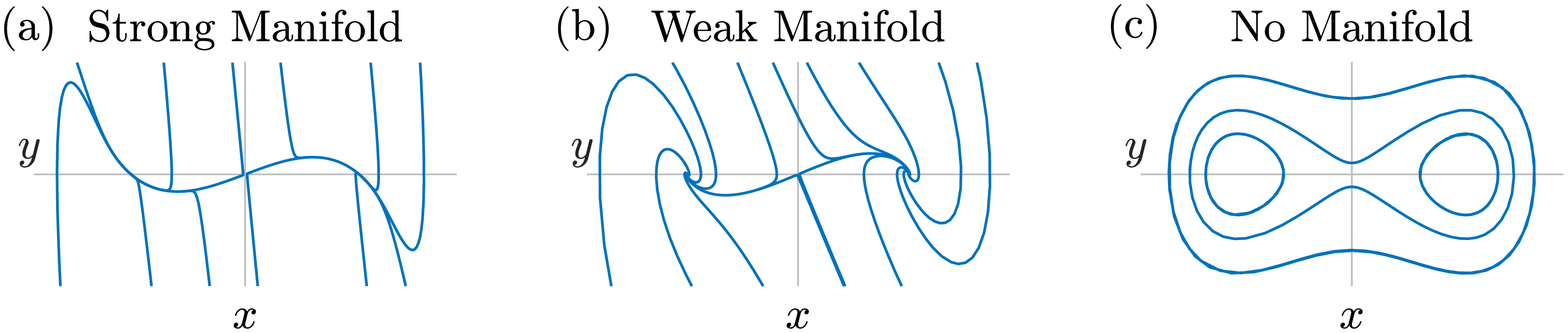}
        \caption{(a) Strong manifold, $\gamma = -5 <<0$. Trajectories are quickly attracted to the strong manifold even far from the fixed points. (b) Weak manifold, $\gamma = -2 <0$. Near $(x,y) = (0,0)$ the manifold is strong and points are attracted to the manifold.  However, away from $(x,y) = (0,0)$, the manifold dissipates and spiral sinks form at the outer fixed points. (c) Without a damping term $\gamma=0$ no manifold exists.}
        \label{fig:manifold}
\end{figure}

Damping in our nonlinear system generates the formation of manifolds or low-dimensional spaces that attract trajectories. If we take $|\gamma|$ to be increasingly large, the vertical line $T = \gamma$ that holds our fixed points in the trace-determinant plane moves away from the origin and any spiral fixed points transition across the curve $T = D^2/4$ becoming nodal fixed points. A polynomial invariant manifold appears, connecting the leading eigenvectors of each fixed point's linear system. We can find this invariant manifold with an asymptotic expansion at the fixed points $y = \sum_{k = 1}^n \alpha_i (x-x^*)^k$. Strong invariant manifolds funnel all points onto the same path making trajectories highly predictable. Manifold strength can be determined analytically for a given system by observing how far below the curve $Det = Tr^2/4$ the fixed points fall in the trace-determinant plane. As points approach this curve the manifold weakens and as they surface above it the manifold dissolves in the given region.
Figure \ref{fig:manifold}(a) shows a heavily damped system in which all fixed points are nodal or saddles, creating a distinct subspace onto which trajectories converge while Figure \ref{fig:manifold}(b) shows the same system but with a weaker damping parameter, the manifold disappears at the outer fixed points which have turned into spirals. Figure \ref{fig:manifold}(c) shows this system with no damping, the system is Hamiltonian.

\subsection*{Nonlinear Control}

\begin{figure}[!ht]
    \centering
    \includegraphics[width=0.9\columnwidth]{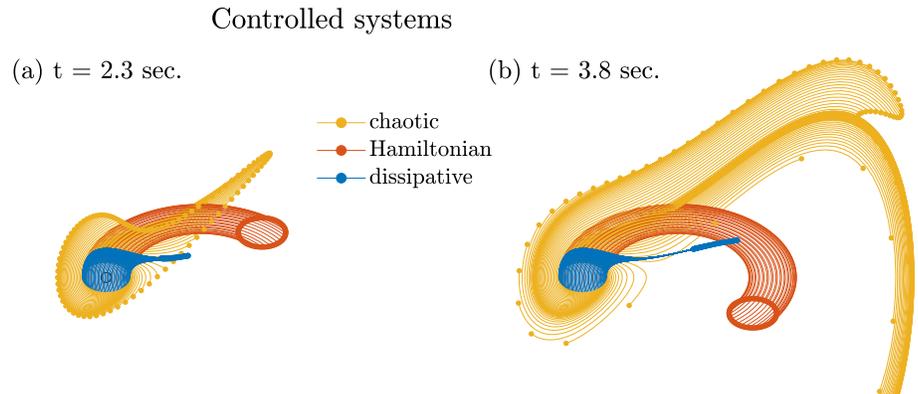}
    \caption{Dynamical system trajectories under the influenced of control signals that make the system chaotic (yellow), Hamiltonian (red), or dissipative (blue). (a) After $t=2.3$ seconds, a small circle of points under chaotic dynamics will spread apart, under Hamiltonian dynamics they maintain the area they encapsulate, and under dissipative dynamics they converge to a low-dimensional subspace. (b) After more time, $t = 3.8$ seconds, the space encapsulated by the chaotic points grows even larger and more deformed, the Hamiltonian space remains constant, and the dissipative system moves along it's subspace until it reaches a fixed point.}
    \label{fig:control_sigs}
\end{figure}

Linear control is a well established area of interest that involves moving a fixed point or stabilizing certain dynamics \cite{sontag_mathematical_2013}. Nonlinear systems can be analyzed and controlled using linear approximations when dynamics are near a fixed point. However, far from fixed points dynamics cannot be presumed to reliably adhere to the linearly approximated dynamics \cite{holmes}. In our work we characterize control methods for moving between fixed points in our nonlinear, cubic polynomial dynamical system.   Specifically, we exclusively consider control that can be achieved with transient control signals. Such signals briefly modify the dynamics, allowing a system to move out of the sphere of influence of a stable fixed point, before allowing the underlying dynamics to dictate dynamical trajectories to one of the fixed points. We consider three types of transient systems (illustrated in Fig.~\ref{fig:control_sigs}) that a control signal can achieve and consider their merits and inadequacies in light of our goal of controlling the system's location over time.\\[.2in]

\noindent {\bf (a) Chaotic systems:}. Initially, we try escaping a stable fixed point and moving to another by changing our system's stable fixed points to unstable fixed points. In our model this is achieved by setting $\gamma >0$. While this modification to the system allows us to leave the fixed point's vicinity, it cannot be used as a control signal as the location of our system under such dynamics becomes unknown. Converting our fixed points to sources turns our system chaotic. A chaotic system cannot reliably move to a desired state and therefore control signals that create this type of uncertainty in the system are inadequate.\\[.2in]

\noindent {\bf (b) Hamiltonian systems:}. A feasible way of moving between fixed points is with a Hamiltonian system. Because we cannot change the stability of the fixed points in our system if we want to maintain control, we must instead eliminate the fixed points. If $\gamma = 0$ our system is Hamiltonian. When we eliminate our local fixed point under this condition, the system leaves the region in a predictable trajectory. While this method of control may bring our system to the locations of the other fixed points in the default system, the termination of the control signal must be precisely timed in order to stay at these fixed points. Hamiltonian systems do not converge and therefore cannot be used for control unless the control signals can be timed well.\\[.2in]

\noindent {\bf (c) Dissipative systems:}.
The ideal way to transition between fixed points is with a dissipative system. Not only are dissipative systems highly predictable, but they contain stable fixed points, allowing controlled systems to converge to a point at or near the target point in the default dynamical system. Our system is dissipative when $\gamma < 0$. If we eliminate a fixed point under these conditions our controlled system will converge to the destination region instead of passing through, as with the Hamiltonian system. With a dissipative system, control can be achieved despite parameter variability, stochasticity, and imprecise control signal timing, making it the objective when designing a controlled system.
\par 
Figure~\ref{fig:control_sigs} shows the behavior of chaotic, Hamiltonian, and dissipative controlled systems over time. In all three cases, points start in a close circular region and then progress in ways stereotypical of their system type. The area encapsulated by the chaotic system's points expands and deforms unpredictably over time. In contrast, the dissipative system's points contract, converging to a discoverable low-dimensional space that they follow to a stable fixed point. The area encapsulated by the Hamiltonian points neither expands nor contracts, keeping its form as the system's points follow their orbital path ad infinitum.

\subsection*{Github repository}
The GitHub repository \textbf{Celegans\_nonlinear\_control} contains code that reproduces select results from this paper and can be found at: \url{https://github.com/mmtree/Celegans_nonlinear_control}. 


\section*{Acknowledgments}
We are deeply indebted to Manuel Zimmer for the use of his \textit{C. elegans} datasets in addition to his extensive expertise and insight that contributed to this project.

\nolinenumbers

%
%
%



\bibliography{Celegans_megan,Zimmer}

\section*{Appendix}

\subsection*{Fixed Points and Stability}

\begin{figure}[!ht]
        \centering
        \includegraphics[width=0.9\columnwidth]{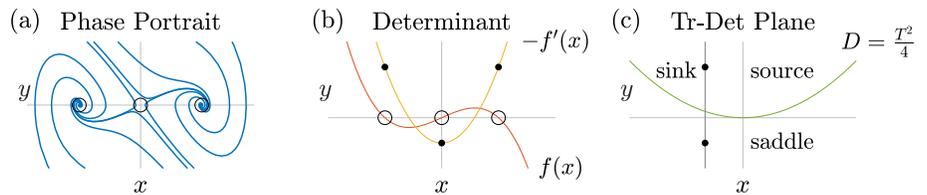}
    \caption{(a) Phase portrait of the polynomial dynamical system $x' = y$, $y' = -(x+1)(x)(x-1) - y$. This system has a saddle at $x^* = 0$ and spiral sinks at $x^* = \pm 1$. (b) The determinant of the Jacobian $Det(J) = -f'(x) = 3x^2-1$ determines fixed point types at $x^*$, the roots of $f(x)=0$. We find that $Det(J^*) = -1,2$. (c) We plot the fixed point trace-determinant values in the trace-determinant plane in order to find their types. All fixed points lie along the vertical line $T =\gamma$. Two of the fixed points are located in the sink region while one point is located in the saddle region of the plane matching what we observe in the phase portrait.}
    \label{fig:TD}
\end{figure}

To find the fixed points of the system we set $(x',y') = (0,0)$ and find that all fixed points lie on the x-axis $y=0$ at the solutions to $f(x) = 0$. In terms of our system parameters the fixed points are denoted by $(x^*,y^*) = (r_i,0)$.
The Jacobian of our system of differential equations tells us how the system behaves at the various fixed points:
\begin{align}
    J = \left[ {\begin{array}{cc}
   0 & 1\\
   f'(x) & \gamma\\
  \end{array} } \right]
\end{align}
The trace and determinant of the Jacobian determine whether fixed points will be sources, sinks, saddles or centers which are mapped in the trace-determinant plane~\cite{holmes,strogatz}. We can evaluate how these types occur and their change in response to the systems parameters $\beta$ and $\gamma$.  Note that
\begin{align}
    &Tr(J) = \gamma\\
    &Det(J) = -f'(x) \, .
\end{align}
Figure~\ref{fig:TD} shows that in the trace-determinant plane all fixed points lie along the vertical line $Tr(J) = \gamma$ located at $Det(J) = -f'(x^*)$. Figure~\ref{fig:TD}(a) shows the phase portrait of a system with two spiral sink fixed points and one saddle. We evaluate the determinant at the fixed points $Det(J^*) = -f(x^*)$, Figure \ref{fig:TD}(b), and map them onto the trace-determinant plane, Figure \ref{fig:TD}(c), showing that our classifications match the fixed point types we observe in the phase portrait.
\par 
Changing parameter values changes our location in the trace-determinant plane. Varying $\gamma$ changes the trace but not the determinant of the system. For $\gamma <0$ all fixed points are either saddles or stable fixed points --- spiral sinks or nodal sinks. However, if $\gamma >0$, all fixed points are either saddles or unstable sources --- nodal or spiral. When $\gamma = 0$ our system only contains saddles and centers. We observe from this that the damping term $\gamma$ alters the stability but not the location of fixed points.
Parameter $a$ does not change the locations of our fixed points but does affect the determinant of our system. Given that $\gamma <0$, changing the sign of $a$ will change our fixed points from stable sinks to unstable saddles and visa-versa. Varying $r_i$ changes both the location and type of fixed points as fixed points can combine, disappear and appear. Because separate parameters control the trace and determinant of the system we can shift them independently.

\subsection*{Approximations around sets of fixed points}

It is well known that nonlinear systems can be approximated with a linear system near fixed points. We extend this technique to approximate nonlinear systems about a number of fixed points with a lower-order system and the dynamics away from all fixed points with only the highest order term.
To approximate a system about a single fixed point $x^*$ we keep the term with this root and make substitutions for all other terms Eq.~\ref{substitute}.
\begin{align}
\label{substitute}
    y'&= a(x-r^*)\prod_{i = 1}^n (r^*-r_i)
\end{align}
The linear approximation is analogous to the Jacobian approximation of the system.
\begin{figure}[!ht]
    \centering
    \begin{tikzpicture}[scale=1]
    \node[inner sep=0pt](russell) at (0,0)
    {\includegraphics[width=0.8\textwidth]{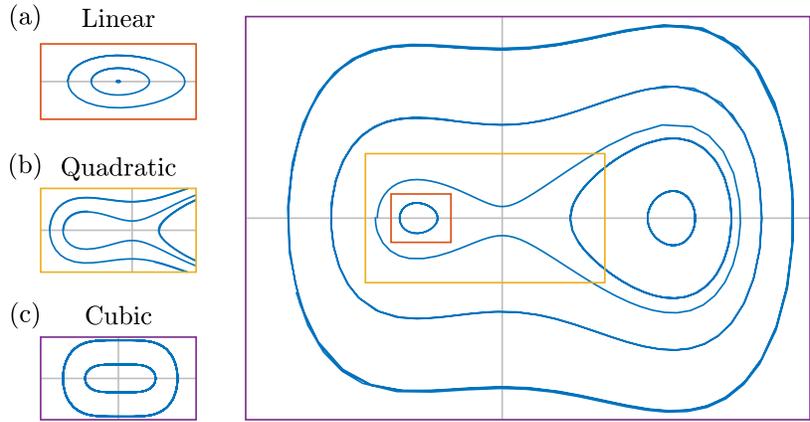}};
    \end{tikzpicture}
    \caption{The dynamics of the polynomial system $x' = y$, $y' = -x(x-1)(x+\frac{1}{2})$ is dominated by different factors depending on the scale and distance from certain fixed points. Near fixed points only the linear terms contribute to the dynamics while far from all fixed points, the leading order cubic terms determines global dynamics. Quadratic terms play a leading role in the intermediate regions. (a) Near the fixed point $x^* = -\frac{1}{2}$ the system can be approximated by a single factor $y' \approx -\frac{3}{4}(x+\frac{1}{2})$. (b) If we zoom out to encapsulate the the region around fixed points $x^* = -\frac{1}{2}$ and $x^* = 0$ but excluding $x^* = 1$, the system can be approximated with two factors corresponding to the enclosed fixed points, $y' \approx x(x+\frac{1}{2})$. (c) If we zoom out even farther to the region enclosing all fixed points we can approximate the system using the highest order term $y' \approx -x^3$.}
    \label{fig:A}
\end{figure}

\subsection*{Global stability}
Stability about individual fixed points do not tell us whether the system is globally stable. Globally unstable systems are undesirable because additional measurements must be taken in order to monitor whether the system is entering a region in which solutions are unbounded and extra control procedures must be established to ensure the system does not enter an unstable region or to bring it out of this region if it does enter. Global stability can be determined by simplifying the dynamical system to the approximate system as $|x|,|y| \rightarrow \infty$. The polynomial in $y'$ can be approximated with its leading order term giving us the following approximate system away from all fixed points.
\begin{align}
    J_{global} = \left[ {\begin{array}{cc}
   0 & 1\\
   a n x^{n-1} & \gamma\\
  \end{array} } \right]
\end{align}
$Det(J_{global})=-anx^{n-1}>0$ if $n \in \mathds{O}, a<0, \gamma <0$ for all $x$ indicating that under these parameter conditions the system is globally stable. If we have a globally stable system we need control regimes only for moving the system out of regions controlled by various fixed points and do not need extra controls to keep the system within stable regions.
\subsection*{System Shifts and Deformations}
\begin{figure}[!htbp]
    \centering
    \includegraphics[width=0.9\columnwidth]{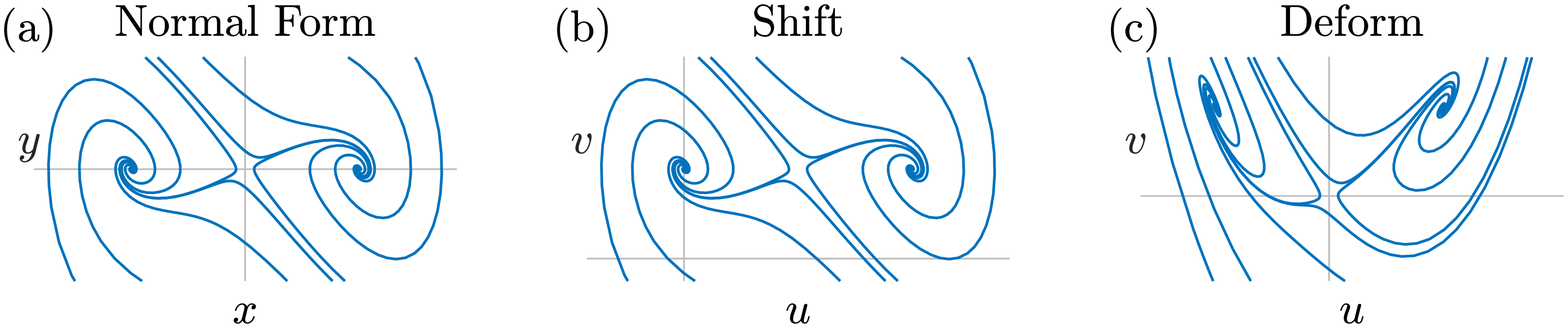}
    \caption{Dynamical system in normal form and two topologically equivalent systems expressing the same qualitative behavior. (a) Normal form expression with dynamics $x' = y$, $y'=-x(x-1)(x+1)-y$ (b) System shifted under the mapping $(u,v) = (x+1, y+1)$. (c) System deformed under the mapping $(u,v) = (x, y+x^2)$.}
    \label{fig:shift_deform}
\end{figure}

The polynomial dynamical systems we have explored are the normal forms of a larger set of nonlinear dynamical systems whose qualitative activity is the same as one of the normal form expressions, yet whose fixed points are not necessarily on the x-axis. These systems are shifts or deformations of the normal form expressions and can be mapped to their corresponding normal form system using a transform a variables that maps all fixed points to the x-axis. The systems activity can be more easily analyzed in the normal form and since the systems are topologically equivalent, all results found in the normal form analysis apply to the original system.  Figure \ref{fig:shift_deform} shows a normal form system as well as two topologically equivalent systems that can be mapped to it through a change of variables. In the first system the system is merely shifted while the second system has undergone a deformation.
Because our characterized collection of normal form systems can be mapped to a large variety of topologically equivalent systems with fixed points anywhere in the $(x,y)$ plane, our set of simple models can be used to understand and model the behavior of systems that express the same qualitative behavior as a normal form system.

\end{document}